\documentclass[12pt]{emulateapj}

\def\lsim{\lower.5ex\hbox{$\; \buildrel < \over \sim \;$}}
\def\gsim{\lower.5ex\hbox{$\; \buildrel > \over \sim \;$}}

\shorttitle{The origin of the 6.4 keV line and H$_2$ ionization in the Galactic center region}
\shortauthors{Dogiel et al.}

\begin{document}


\title{The origin of the 6.4 keV line emission and H$_2$
ionization in the diffuse molecular gas of the Galactic center
region}

\author{V. A. Dogiel$^{1,2}$, D. O. Chernyshov$^1$,
V. Tatischeff$^3$, K.-S. Cheng$^2$, R. Terrier$^4$}

 \affil{$^1$I.E.Tamm Theoretical Physics Division
of P.N.Lebedev Institute of Physics, Leninskii pr. 53, 119991
Moscow, Russia} \affil{$^2$Department of Physics, University of
Hong Kong, Pokfulam Road, Hong Kong, China} \affil{$^3$Centre de
Spectrom\'etrie Nucl\'eaire et de Spectrom\'etrie de Masse,
IN2P3/CNRS and Univ Paris-Sud, 91405 Orsay Campus, France}
\affil{$^4$Astroparticule et Cosmologie, Universit\'e Paris7/CNRS/CEA,
Batiment Condorcet, 75013 Paris, France}

\begin{abstract}
We investigate the origin of the diffuse 6.4~keV line emission
recently detected by {\it Suzaku} and the source of H$_2$
ionization in the diffuse molecular gas of the Galactic center
(GC) region. We show that Fe atoms and H$_2$ molecules in the
diffuse interstellar medium of the GC are not ionized by the same
particles. The Fe atoms are most likely ionized by X-ray photons
emitted by Sgr~A$^\ast$ during a previous period of flaring
activity of the supermassive black hole. The measured longitudinal
intensity distribution of the diffuse 6.4~keV line emission is best explained
if the past activity of Sgr~A$^\ast$ lasted at least several hundred years
and released a mean $2-100$~keV luminosity $\gsim 10^{38}$~erg~s$^{-1}$.
The H$_2$ molecules of the diffuse
gas can not be ionized by photons from Sgr A$^\ast$, because soft
photons are strongly absorbed in the interstellar gas around the
central black hole. The molecular hydrogen in the GC region is
most likely ionized by low-energy cosmic rays, probably protons
rather than electrons, whose contribution into the diffuse 6.4 keV
line emission is negligible.
\end{abstract}



%

\section{Introduction}

The Central Molecular Zone (CMZ) has long been known as a thin
layer of about $300\times 50$~pc in size, containing a total of
$2.4\times 10^7$~$M_\odot$ of dense ($n\sim 10^4$ cm$^{-3}$), high
filling factor ($f>0.1$) molecular material orbiting the Galactic
center \citep[see the review of][]{katia07,katia12}. This
canonical picture was drastically changed after the discovery of
H$_3^+$ absorption lines generated by ionization of H$_2$
molecules. Observations of \citet{mccall02} and \citet{oka05}
showed that H$_3^+$ are mainly generated in diffuse ($n\sim 10^2$
cm$^{-3}$) clouds, where the ratio of H$_3^+$ to molecular
hydrogen abundance is 10 times higher than in dense clouds. This
diffuse gas has a high filling factor ($f_V \sim 0.3$), an
unusually high temperature ($T_H\sim 250$~K) and a high and almost
uniform ionization rate ($\zeta_2 \sim (1-3)\times 10^{-15}$
s$^{-1}$) throughout the CMZ \citep[see][and references
therein]{geb12}. The source of this gas ionization is still
debated.

Neutral Fe K$\alpha$ line emission at 6.4~keV is observed from
several dense  Galactic Center (GC) molecular clouds.
 Detections of time variability of the X-ray emission from Sgr~B2
\citep[see e.g.][]{nob11} and from several clouds within $15'$ to
the east of Sgr~A$^\ast$ \citep{mun07,ponti10}, strongly suggest
that the Fe K$\alpha$ line emission from these regions is a
fluorescence radiation excited by a past X-ray flare from the
supermassive black hole. In this model, the variability of the
line flux results from the propagation of an X-ray light front
emitted by Sgr~A$^\ast$ more than $\sim 100$~years ago.

The neutral Fe K$\alpha$ line can also be generated by charged
particles \citep[see][]{dog98,valinia}, and observations do not
exclude that the 6.4 keV emission from some of GC clouds is
produced by CR electrons or protons and nuclei \citep[see,
e.g.,][]{fuk09,capelli11,tati12,yus1,yus3,dog09,dog11}.

 In addition to the bright X-ray emission from dense clouds,
\citet{uchiyama2012} recently found with {\it Suzaku} diffuse
emission at 6.4 keV from an extended region of the GC region, with
a scale length in longitude of $\ell\sim 0.6^\circ$  and an extent
in latitude of $b\sim 0.2^\circ$--$0.4^\circ$  when all known
contributions from point sources and compact clouds ($n>> 10^2$
cm$^{-3}$) were subtracted. These authors concluded that this
emission  is truly diffuse. Based on the equivalent width (EW) of
the line ($\sim 460$ eV), they also concluded that the origin of
the Fe I K$\alpha$ line emission from the diffuse gas in the GC
might be different from that of the dense clouds. However,
\citet{hea13} recently suggested that unresolved stellar sources
may make an important contribution to the observed diffuse
emission at 6.4~keV.

We investigate in this paper if the diffuse 6.4~keV line emission
 and the H$_3^+$ absorption lines from  the diffuse
molecular gas  ($n\sim 10^2$ cm$^{-3}$, i.e. outside dense clouds)
can have the same origin, i.e. if Fe atoms and H$_2$ molecules in
this medium can be ionized by the same particles. In this
investigation, we obtain new constraints on the past X-ray flaring
activity of Sgr~A$^\ast$, as well as on the density of low-energy
cosmic rays (LECRs) in the GC region.

\section{H$_2$ ionization and 6.4~keV line emission from cosmic rays}

The ratio of the number of 6.4~keV photons and H$_2$ ionization
produced by CRs of type $i$ (electrons or protons)
propagating in diffuse molecular gas can be estimated from
\begin{equation}
X_{6.4,i}\approx{\eta_{\rm Fe} \over f_{\rm H_2}}{\int_{I({\rm
Fe~K})}^{E_{\rm max}}N_i(E)\sigma_{i{\rm Fe}}^{\rm
K\alpha}(E)v_i(E)dE \over \int_{I({\rm H_2})}^{E_{\rm
max}}N_i(E)\sigma_{i{\rm H_2}}^{\rm ioni}(E)v_i(E)dE}~,
\label{eq:x64_3}
\end{equation}
where $I({\rm Fe~K})=7.1$~keV and $I({\rm H_2})=15.6$~eV are the
ionization potentials of the K shell of Fe and of H$_2$,
respectively, $\eta_{\rm Fe}$ is the Fe abundance, $f_{\rm
H_2}=n({\rm H_2})/(2n({\rm H_2})+n({\rm H}))$ the fractional
density of H$_2$ molecules relative to the total number of H
atoms, $N_i(E)$ the differential  equilibrium number of
CRs of type $i$ propagating in the diffuse gas, $v_i(E)$
the velocity of these particles, $E_{\rm max}$ their maximum
kinetic energy, $\sigma_{i{\rm Fe}}^{\rm K\alpha}(E)$ the cross
section for producing the 6.4~keV line by interaction of fast
particles of type $i$ with Fe atoms, and $\sigma_{i{\rm H_2}}^{\rm
ioni}(E)$ the cross section for ionization of the H$_2$ molecule
by the particle $i$. We have neglected in this equation the
contribution of secondary electrons for both 6.4~keV line
production and H$_2$ ionization.

\begin{figure}
\centering
\includegraphics[width=0.4\textwidth]{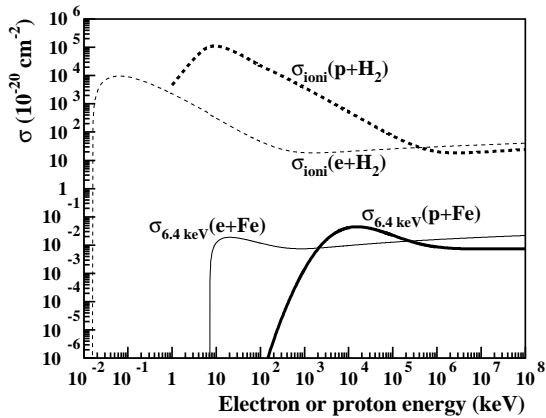}
\caption{Cross sections involved in the calculation of the 6.4~keV
photon yield per H$_2$ ionization in the CR model. {\it
Solid lines}: cross sections for producing the 6.4 keV line by
interaction of fast electrons ({\it thin line}) and protons
({\it thick line}) with Fe atoms. {\it Dashed lines}: H$_2$
ionization cross sections for impact of electrons ({\it thin line})
and protons ({\it thick line}).} \label{fig:x64sig}
\end{figure}

The cross sections $\sigma_{i{\rm Fe}}^{\rm K\alpha}$ and
$\sigma_{i{\rm H_2}}^{\rm ioni}$ are shown in
Fig.~\ref{fig:x64sig} for both CR electrons and protons.
The H$_2$ ionization cross sections were taken from \citet{pad09}.
In addition to the CR impact ionization reaction $i+{\rm
H_2} \rightarrow i+{\rm H_2}^+ + e^-$, the proton cross section
$\sigma_{p{\rm H_2}}^{\rm ioni}$ includes the charge-exchange
process $p+{\rm H_2} \rightarrow {\rm H}+{\rm H_2}^+$, which is
dominant below $\sim 40$~keV \citep[see][Figure~1]{pad09}. The
cross sections $\sigma_{e{\rm Fe}}^{\rm K\alpha}$ and
$\sigma_{p{\rm Fe}}^{\rm K\alpha}$ were calculated as in
\citet{tati12}. We see that the H$_2$ ionization cross sections
are much higher than that for the X-ray line production. At
relativistic energies, the difference is by a factor $\sim 2000$
for electrons and $\sim 3000$ for protons.

\begin{figure}
\centering
\includegraphics[width=0.4\textwidth]{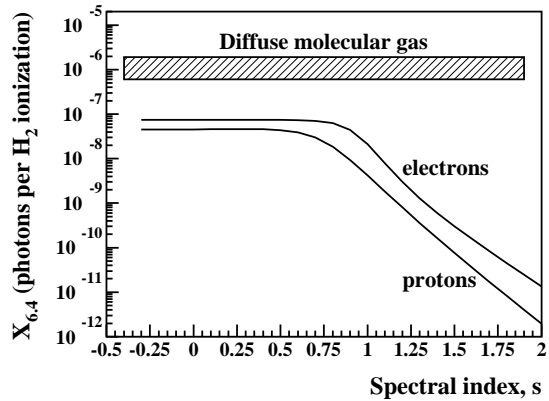}
\caption{6.4~keV photon yield per H$_2$ ionization in diffuse
molecular gas. Theoretical values obtained from
Eq.~(\ref{eq:x64_3}) are compared with the range deduced from
X-ray and H$_3^+$ observations, assuming  $\langle{n_{\rm
H_2}}\rangle=50$~cm$^{-3}$ and $R=200$~pc (Eq.~(\ref{eq:x64_2})).
We used for the calculations $\eta_{\rm Fe}=6.9\times 10^{-5}$,
which is twice the solar abundance \citep{asp09}, and $f_{\rm
H_2}=0.5$, i.e. $n({\rm H_2})\gg n({\rm H})$.}
\label{fig:x64}
\end{figure}

The propagated CR spectrum $N_i(E)$ could be calculated in
the framework of a given model for the source of these particles.
Here instead, for the sake of generality, we use a simple power
law in kinetic energy, $N_i(E) \propto E^{-s}$, allowing the
spectral index $s$ to take any value within a reasonable range.
Calculated values of $X_{6.4,i}$ are shown in Fig.~\ref{fig:x64}
as a function of $s$.

Also shown in this Figure is the 6.4~keV photon yield
per H$_2$ ionization deduced from observations. The latter is
estimated by assuming that the diffuse 6.4~keV line emission
is emitted from a thick disk centered on Sgr~A$^\ast$, with
volume $V=\pi R^2h$ and solid angle at the observer position
$\Omega=2Rh/D^2$ ($R$ and $h$ are the
disk radius and height perpendicular to the Galactic plane,
respectively, and $D$ the distance to the Galactic center):
\begin{equation}
X_{6.4}={4 \pi D^2  I_{6.4}  \Omega   \over V \langle{n_{\rm
H_2}}\rangle \zeta_2}~.
\label{eq:x64_1}
\end{equation}
Here, $I_{6.4}=(7.3\pm 0.7)$~ph~cm$^{-2}$~s$^{-1}$~sr$^{-1}$ is the
measured intensity of the diffuse Fe K$\alpha$ line emission near
Sgr~A$^\ast$ \citep{uchiyama2012}, $\zeta_2 \approx (1-3)\times 10^{-15}$~s$^{-1}$
is the estimated ionization rate of the H$_2$ molecule in the diffuse
molecular gas \citep{goto08}, and $\langle{n_{\rm H_2}}\rangle$ the
mean density in the disk of H$_2$ molecules of the diffuse gas
($\langle{n_{\rm H_2}}\rangle=f_Vn_{\rm H_2}$, where $f_V$ is
the volume filling factor of the diffuse molecular gas and
$n_{\rm H_2}$ the H$_2$ number density in this gaseous component).
We then get
\begin{eqnarray}
&&X_{6.4}={8 I_{6.4} \over \langle{n_{\rm H_2}}\rangle \zeta_2 R} = \label{eq:x64_2}\\
&& \approx (0.6 - 1.9)\times 10^{-6} \times
\bigg({\langle{n_{\rm H_2}}\rangle \over 50~{\rm cm}^{-3}}
\bigg)^{-1}\bigg({R \over 200~{\rm~pc}}\bigg)^{-1}~.\nonumber
\end{eqnarray}

We see in Fig.~\ref{fig:x64} that the theoretical photon yield
is almost constant for low values of $s$, which reflects the
constancy of the cross section ratio $\sigma_{i{\rm Fe}}^{\rm
K\alpha}/\sigma_{i{\rm H_2}}^{\rm ioni}$ at relativistic energies
(Fig.~\ref{fig:x64sig}). The calculated photon yield amounts to
$\lsim 10$\% of the measured value at low $s$ and rapidly declines
for $s \ga 0.75$.

 \citet{yus3} estimated that a population of LECR
electrons responsible for the ionization of the diffuse H$_2$ gas
would contribute to  $\sim 10$\% of the diffuse 6.4~keV line emission
detected by {\it Suzaku}, which is consistent with the maximum
contribution we expect for a hard electron spectrum (see Fig.~\ref{fig:x64}).
These authors also suggested that the remaining 90\% of the X-ray
emission is produced by interactions of electrons with a denser
($n\sim 10^3$~cm$^{-3}$) molecular gas. But this would imply the
existence of a massive ($\sim 10^7~M_\odot$) molecular gas component
with a high ionization rate ($\zeta_2 \gsim 10^{-14}$~s$^{-1}$),
which should have been detected with H$_3^+$ observations
\citep[see, e.g.,][]{geb12}. Independent of the exact density of
the diffuse molecular gas, the comparison of the cross sections
$\sigma_{i{\rm Fe}}^{\rm K\alpha}$ and $\sigma_{i{\rm H_2}}^{\rm ioni}$
(Fig.~\ref{fig:x64sig}) show that the bulk of the diffuse 6.4~keV line
emission is not produced by CRs.

\section{The origin of the diffuse 6.4~keV line emission: past flaring activity of Sgr~A$^\ast$}

Unlike the 6.4 keV line emission from dense clouds, the X-ray
fluorescence emission from the diffuse gas is predicted to be
almost constant for several hundred years, which is the time
needed for a photon to cross the CMZ \citep[see][]{chern12}. We
use here for the gaseous disk radius $R=200$~pc.

The expected flux in the 6.4 keV line depends on two parameters of
the flare from Sgr~A$^\ast$: its luminosity $L_X$ and duration
$\Delta t$. The flare duration estimated from observations ranges
from $\sim 10$~yr \citep[see][]{yu11} to 500~yr \citep[][]{ryu12}.
The flare probably ended at $T \sim 70-150$~years ago \citep[see
e.g.][]{ponti10,ryu12}. The required luminosity $L_X$ in a given
energy range $E_{\rm min}$~--~$E_{\rm max}$ can be estimated for
each value of $\Delta t$ from the observed intensity of diffuse
6.4~keV line emission.

As \citet{koya2} and \citet{terrier} showed, the
differential spectrum of primary X-ray photons emitted by Sgr~A$^\ast$ can be
described by a power-law in the $2-100$~keV energy range:
\begin{equation}
n_{\rm ph}(E_X,r=0)\propto E_X^{-2}~. \label{lsp}
\end{equation}
The distribution of primary photons in the disk as a function of
energy $E_X$ and radius $r$ is then given by
\begin{eqnarray}
n_{\rm ph}(E_X,r) &=& \frac{Q_0}{4\pi c
r^2E_X^2}\exp\big(- \langle{n_{\rm
H_2}}\rangle \sigma_{\rm abs}(E_X)r/f_{\rm H_2}\big)\times \nonumber \\
& & \theta(r-cT)\theta(cT+c\Delta t-r)~,
\label{phs}
\end{eqnarray}
where $\theta(x)$ is the Heaviside function, $\sigma_{\rm
abs}$ the photoelectric absorption cross section per H atom
\citep{bal92}, $c$ the speed of light, and $Q_0$ a normalization
constant derived from the estimated luminosity of Sgr A$^*$ in the
energy range $E_{\rm min}-E_{\rm max}$ (below we use the range
$2-100$~keV):
\begin{equation}
L_X=Q_0\ln \left(E_{\rm max}/E_{\rm min}\right)~.
\end{equation}

A distant observer sees at present the reflection of a light front
emitted by Sgr~A$^\ast$ at time $t$ in the past as a parabola
\citep[see, e.g.,][]{SCh98},
\begin{equation}
\frac{z}{c} =\frac{1}{2t}\left[t^2 -
\left(\frac{x}{c}\right)^2\right]\,, \label{zc}
\end{equation}
where the coordinates $x \simeq \ell D$ and $z$ are perpendicular and
along the line of sight, respectively. The observable longitudinal
distribution of 6.4~keV line emission is then given by
\begin{eqnarray}
I_{6.4}(\ell) &=& {\eta _{\rm Fe} \langle{n_{\rm H_2}}\rangle
\over 4\pi f_{\rm H_2}} \int_{I({\rm Fe~K})}^{E_{\rm max}}
\sigma_{X{\rm Fe}}^{\rm K\alpha}(E_X) \times \nonumber \\
& & \big( \int_{z_1}^{z_2} n_{\rm ph}[E_X,\sqrt{(\ell D)^2+z^2}]
dz \big) dE_X~, \label{i6.4}
\end{eqnarray}
where $z_1$ and $z_2$ are calculated from Eq.~(\ref{zc}) for $t=T$ and $t=T+\Delta t$,
respectively, and $\sigma_{X{\rm Fe}}^{\rm K\alpha}$ is the cross section for
producing the 6.4~keV line by Fe K-shell photoionization \citep[see][]{tati03}.

\begin{figure}
\centering
\includegraphics[width=0.45\textwidth]{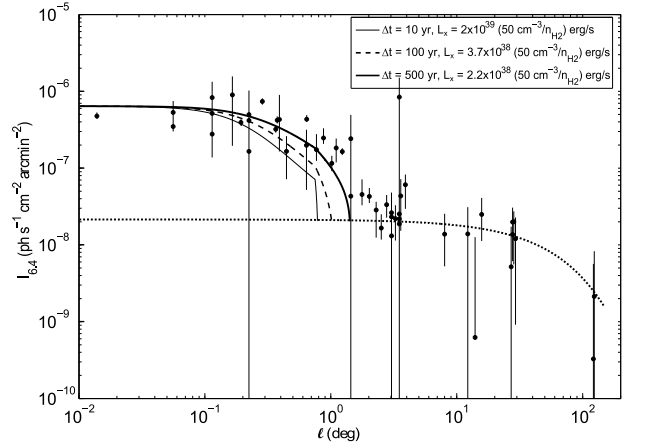}
\caption{Longitudinal distribution of the 6.4~keV line generated
by primary photons in the GC region calculated  for the Fe
abundance $\eta_{\rm Fe}=6.9\times 10^{-5}$, which is twice the
solar abundance. The data points are from \citet{uchiyama2012}.
The modeled emission from the Galactic ridge ({\it dotted line}) is
also from that paper.}\label{xrn}
\end{figure}

The results obtained for different values of $\Delta t$ and $L_X$
are shown in Fig.~\ref{xrn}, together with the {\it Suzaku} data
\citep{uchiyama2012}. The amount of illuminated diffuse gas
depends on the flare duration. For $\Delta t=10$~yr, the required
luminosity of Sgr~A$^\ast$ is about $10^{39}$~erg~s$^{-1}$. But
for $\Delta t \ga 350$~yr most of the disk emits the Fe K$\alpha$
line, whose brightness is then independent of the flare duration.
The required X-ray luminosity in this case is about
$10^{38}$~erg~s$^{-1}$. We see in Fig.~\ref{xrn} that the {\it
Suzaku} data are in better agreement with the assumption of a long
flare duration.  The possibility of such a long period of activity
with some sporadic flux variability is not excluded by
observations of the Sgr~C region \citep{ryu12}. However, Chandra
observations of clouds within the central 30~pc reveal they are
probably illuminated by two short flares \citep{clavel13}.

We also see in Fig.~\ref{xrn} that even for $\Delta
t=500$~yr, the model slightly underestimates the measured 6.4~keV
line flux at longitudinal distances from Sgr~A$^\ast$ of $\sim
1^\circ$. This suggests that the warm and diffuse gas extends
beyond the CMZ, which might be already suggested by the measured
outwards movement of this gas component \citep[see][]{geb12}.

\section{The origin of the H$_2$ ionization in the diffuse molecular gas:
X-ray photons vs LECRs}

If the flare ended $\sim 100$~yr ago, then soft photons emitted by
Sgr A* were absorbed in the dense interstellar medium of the GC.
For the hydrogen density $\sim 100$~cm$^{-3}$, photons with
energies $E_X > 1$ keV survive in this region. From the Compton
echo, we know the spectral shape of primary photons of $E_X>1$ keV
emitted by Sgr A*: it is presented by Eq. (\ref{lsp}).

Because the ionization rate $\zeta_2$ is expected to be
strongly non-uniform, we calculate instead the H$_3^+$ column
density, $N({\rm H_3^+})$,  which at the longitudinal distance
$\ell$ from Sgr A$^\ast$ has the form
\begin{equation}
N({\rm H_3^+})=\int\limits_{l} n_{{\rm H}_3^+}(r, \tilde{t}) dl~.
\label{h3+}
\end{equation}
Here, $l$ is a trajectory of an IR photon through the disk diffuse
gas and $n_{{\rm H}_3^+}(r, \tilde{t})$ is the H$_3^+$ density in
the disk as a function of radius $r$ and time $\tilde{t}$, where
the tilde indicates that IR photons crossing the gaseous disk
interact with H$_3^+$ ions of different ``ages''.  Therefore, the
geometrical formalism entering in the calculation of $N({\rm
H_3^+})$ (Eq.~\ref{h3+}) is similar to that applied for
$I_{6.4}(\ell)$ (Eq.~\ref{i6.4}).

The process of H$_2$ photoionization leading to H$_3^+$ production
is time variable \citep[see, e.g.,][]{goto13}:
\begin{equation}
\frac{\partial n_{{\rm H}_3^+}}{\partial t}  = \zeta_2(r,t) \langle n_{\rm H_2}\rangle - \langle
v_e\sigma_{{\rm H_3^+}}^{\rm rec} \rangle n_e n_{{\rm H}_3^+}\,,
\label{nh3+}
\end{equation}
where $\langle v_e\sigma_{{\rm H_3^+}}^{\rm rec} \rangle$ is the rate
of H$_3^+$ dissociative recombination and $n_e$ the density of free electrons.
The ionization rate is due to both photoelectric ionization and Compton
scattering:
\begin{eqnarray}
& & \zeta_2(r,t) \simeq \int\limits_{I({\rm H_2})}^{E_{\rm max}}dE_X \sigma_{X{\rm
H_2}}^{\rm ioni}(E_X) c n_{\rm ph}(E_X,r,t) M_{\rm sec}(E_X) + \nonumber \\
& & 2\int\limits_{E_{\rm I}}^{E_{\rm max}}dE_X cn_{\rm ph}(E_X,r,t)
\int \limits_{I({\rm H_2})}^{E^{\rm max}_e} dE_e \frac{d\sigma_c}{dE_e}
M_{\rm sec}(E_e)~,
\end{eqnarray}
where $\sigma_{X{\rm H_2}}^{\rm ioni}$ is the H$_2$ photoelectric
ionization cross section \citep{yan}, $d\sigma_c/dE_e$ the
Klein-Nishina differential cross-section as a function of the
energy of the recoil electron $E_e$, $M_{\rm
sec}(E_e)=[E_e-I({\rm H_2})]/W$ with $W \approx 40$~eV
\citep[see][]{dalg99} the mean multiplicity of H$_2$ ionization
by a secondary electron, $E_{\rm I} \approx \sqrt{m_ec^2{\rm
I({\rm H_2})/2}}$ the minimum energy of an X-ray photon to
ionize an H$_2$ molecule by Compton scattering, $E^{\rm max}_e =
2E_X^2/(m_ec^2 + 2E_X)$, and $n_{\rm ph}(E_X,r,t)$ is given by
Eq.~(\ref{phs}).

The fraction of free electron can be approximated by the abundance
of  singly ionized carbon, C$^+$, assuming that nearly all free
electrons are the result of C photoionization \citep[see
e.g.][]{oka06}. Here, the C abundance is taken to be twice that measured by
\citet{sofia04} in the Galactic disk, i.e. $\eta_{\rm C} = 3.2\times 10^{-4}$.

The H$_3^+$ recombination rate is given in
\citet{mccall04} as a function of the gas temperature $T_{\rm H}$ in K:
 \begin{equation}
 \langle v_e\sigma_{X{\rm H_3^+}}^{\rm rec} \rangle = -1.3\times 10^{-8}+1.27\times 10^{-6}T_{\rm H}^{-0.48}~.
 \label{r_rate}
 \end{equation}
For $T_{\rm H}\sim 250$~K, the rate is $\sim 8 \times 10^{-8}$~cm$^3$~s$^{-1}$
and the characteristic recombination time $t_{\rm rec}=[n_e\langle
v_e\sigma_{{\rm H_3^+}}^{\rm rec} \rangle]^{-1}\sim 10$~yr, which is
much smaller than the light propagation time across the CMZ.

The solution of Eq.~(\ref{nh3+}) takes the form
\begin{eqnarray}
n({\rm H}_3^+) & = & \int\limits_0^t d\tau
\exp\left[ \langle v_e\sigma_{{\rm H_3^+}}^{\rm rec} \rangle n_e (\tau-t) \right] \zeta_2(r,\tau)
\langle n_{{\rm H}_2} \rangle~. \nonumber \\
&&
\end{eqnarray}

\begin{figure}
\centering
\includegraphics[width=0.45\textwidth]{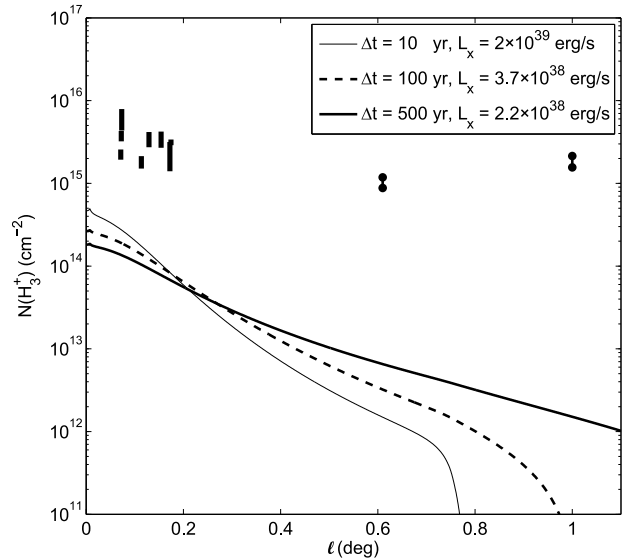} \caption{${\rm H_3^+}$
column density provided by primary photons from Sgr~A$^\ast$, for
three values of $(\Delta t,L_X)$. The data points are from
\citet{goto08} and \citet{goto11}.} \label{nh3}
\end{figure}

Absorption of low-energy photons in the dense material surrounding
Sgr~A$^\ast$  \citep[see][]{katia12} can be essential. However, to
derive an upper limit on $N({\rm H_3^+})$, we neglect here the
photoabsorption in this dense medium and consider only that
occurring in the diffuse molecular gas of mean density $\langle
n_{\rm H_2} \rangle =50$~cm$^{-3}$. H$_3^+$ column densities
calculated for $T=100$~yr and for three values of $(\Delta t,L_X)$
are shown in Fig. \ref{nh3}. We see that $N({\rm H_3^+})$  is no
more than $4\times 10^{14}$~cm$^{-2}$, which is less than the
observed values, which lie in the range
$10^{15}$--$10^{16}$~cm$^{-2}$  \citep[see][]{goto08,goto11}.
Thus, although flare photons from Sgr~A$^\ast$ probably generate
most of the 6.4~keV line emission from the GC region, they are
most likely not the main source of H$_2$ ionization in the diffuse
gas.

Ionization of molecular hydrogen can be effectively produced by
LECRs. \citet{yus3} suggest that the necessary
ionization rate, $\zeta_2 \sim (1-3)\times 10^{-15}$~s$^{-1}$, can be
provided by low-energy electrons ($E\ga 100$ keV). These authors
estimated the energy distribution of radioemitting electrons ($E>1$ GeV)
in the GC region assuming for the mean magnetic field strength
$B \sim 10^{-5}$~G, and then extrapolated the derived electron spectrum
to lower energies. However, 100-keV electrons have a short lifetime
in the diffuse molecular gas ($\sim 100$~yr), which makes it difficult
to explain the measured uniformity of the H$_2$ ionization rate, unless
low-energy electrons are constantly produced all over the disk.
Besides, if the magnetic field strength in the GC amounts to
$B\ga 5\times 10^{-5}$~G as derived by \citet{crock10}, then the
calculations of \citet{yus3} would give $\zeta_2 \la  10^{-16}$ s$^{-1}$.
We also note that the inclusion of Coulomb energy losses in these
calculations would flatten the spectrum of radioemitting electrons
at low energies, thus reducing the ionization rate.

For all of these reasons, molecular hydrogen in the diffuse gas is
more likely to be ionized by subrelativistic protons. These
particles can be generated either by star accretion onto the
central black hole \citep[][]{dog09b}, or by diffusive shock
acceleration in supernova remnants (SNRs). The latter supply in
the GC region a total kinetic power of $\sim 10^{40}$~erg~s$^{-1}$
\citep[see][]{crock11}. In comparison, the proton power needed to
explain the H$_2$ ionization rate is
$\dot{W}_p \sim \zeta_2 \langle n_{\rm H_2} \rangle V W \sim 2 \times 10^{39}$~erg~s$^{-1}$,
such that an efficiency of proton acceleration in SNRs of
$\sim 20\%$ could account for the H$_3^+$ line measurements.

\section{Conclusions}

We have shown that the diffuse 6.4~keV line radiation from the GC
region is most likely produced by the hard X-ray photon emission
from Sgr~A$^\ast$ that also produces fluorescence X-ray radiation
in several dense molecular clouds. The longitudinal intensity
distribution of the diffuse Fe K$\alpha$ line emission thus
provides an additional  constrain on the past activity of the
central black hole. Our results on the past X-ray luminosity of
Sgr~A$^\ast$ are broadly consistent with that obtained from the
6.4~keV line radiation of dense clouds by \citet{capelli12},
$L_X(2-10~{\rm keV})\sim 10^{38}$~erg~s$^{-1}$ if the flare
duration was about 100~yr, and by \citet{ponti10}, $L_X(2-100~{\rm
keV})\sim 10^{39}$~erg~s$^{-1}$ if $\Delta t \sim 10$~yr. But the
measured distribution of the diffuse 6.4~keV line emission
strongly suggests that the past activity of Sgr~A$^\ast$ lasted at
least several hundred years. {\it Suzaku} observations of the
Sgr~C molecular cloud complex does not exclude also that
Sgr~A$^\ast$ was continuously active with sporadic flux
variabilities in the past 50 to 500 years \citep{ryu12}. However,
the overall agreement of our results on $L_X$ with that previously
obtained from the X-ray emission of dense molecular clouds
suggests that most of the large-scale 6.4~keV line emission from
the GC region is truly diffuse and not due to a collection of
unresolved point sources.

On the other hand, high-energy photons emitted by Sgr~A$^\ast$ are not
responsible for the ionization of the diffuse molecular gas. The H$_2$
molecules in this gas are very likely ionized by LECRs,
probably protons accelerated in SNRs, whose contribution into the diffuse 6.4~keV
line emission is negligible.

\acknowledgments We are very grateful to Katia Ferri\`{e}re,
Masayoshi Nobukawa, Takeshi Oka and Bob Warwick for their very
useful comments and critical reading of the text, and to Miwa Goto
and co-authors who sent us their paper before publication. VAD,
DOC, VT and RT acknowledge support from the International Space
Science Institute to the International Team 216. VAD and DOC are
 supported by the RFFI grant 12-02-00005-a. DOC is also
supported in parts by the RFFI grant 12-02-31648 and the LPI
Educational-Scientific Complex. KSC is supported by a grant under
HKU 2011/10p.

\end{document}